\documentclass[conference]{IEEEtran}
\IEEEoverridecommandlockouts
\usepackage{cite}
\usepackage{amsmath,amssymb,amsfonts}
\usepackage{algorithmic}
\usepackage{graphicx}
\usepackage{textcomp}
\usepackage{xcolor}
\usepackage{comment}
\usepackage{multirow}
\usepackage{array}
\usepackage{listings}

\newcolumntype{P}[1]{>{\centering\arraybackslash}p{#1}}
\renewcommand*{\arraystretch}{1.2}
\newcolumntype{M}[1]{>{\centering\arraybackslash}m{#1}}
\def\BibTeX{{\rm B\kern-.05em{\sc i\kern-.025em b}\kern-.08em
    T\kern-.1667em\lower.7ex\hbox{E}\kern-.125emX}}
\usepackage{hyperref}
\usepackage{tikz}
\usetikzlibrary{shapes}
\usepackage{xcolor}
\newlength\MAX  

\usepackage{subcaption}
\usepackage{graphicx}
\usepackage{tabularx}
\usepackage{listings}
\usepackage{amsmath}  
\usepackage{array}
\usepackage{booktabs}
\usepackage{titlesec}

\titlespacing{\section}{0pt}{0.3ex plus 0.2ex minus .2ex}{0.3ex}
\titlespacing{\subsection}{0pt}{0.2ex plus 0.13ex minus .1ex}{0.1ex}

\begin{document}
 \makeatletter
\newcommand{\linebreakand}{%
\end{@IEEEauthorhalign}
\hfill\mbox{}\par
\mbox{}\hfill\begin{@IEEEauthorhalign}
}
\makeatother
\title{From Student Risk Prediction to SC$^2$R:\\[-0.2em] Semantics-Constrained Counterfactual Recourse for\\[-0.2em] Educational Decision Support\vspace{-0.5cm}}

\author{
	\IEEEauthorblockN{Ngoc Luyen Le\IEEEauthorrefmark{1}\IEEEauthorrefmark{2}\\[-0.2em]\textit{ngoc-luyen.le@hds.utc.fr}\vspace{-0.3cm}}
	\and
	\IEEEauthorblockN{Marie-Hélène Abel\IEEEauthorrefmark{2}\\[-0.2em]\textit{marie-helene.abel@hds.utc.fr}\vspace{-0.3cm}}
	\and
	\IEEEauthorblockN{Bertrand Laforge\IEEEauthorrefmark{3}\\[-0.2em]\textit{laforge@lpnhe.in2p3.fr}\vspace{-0.3cm}}

	\linebreakand
	\IEEEauthorblockA{
		\IEEEauthorrefmark{1}Gamaizer, 93340 Le Raincy, France.
		\\[-0.2em]
		\IEEEauthorrefmark{2}Université de Technologie de Compiègne, CNRS, Heudiasyc (Heuristics and Diagnosis of Complex Systems),\\[-0.2em] CS 60319 - 60203 Compiègne Cedex, France.
			\\[-0.2em]
		\IEEEauthorrefmark{3}Sorbonne Universit\'{e}, CNRS UMR 7585, LPMHE (Laboratoire de Physique Nucl\'{e}aire et des Hautes \'{E}nergies),\\[-0.2em] 75252 Paris cedex 05, France
		\vspace{-0.8cm}
	}
}
\maketitle
\begin{abstract}
	Learning analytics models can identify students at risk of poor performance, but they do not directly indicate which interventions are feasible, actionable, and compatible with educational constraints. This paper introduces \textbf{SC$^2$R}, a semantics-constrained counterfactual recourse framework for educational decision support. \textbf{SC$^2$R} combines a calibrated predictive model, integer-programming-based recourse generation over discrete action variables, a lightweight RDF vocabulary for intervention-plan representation, and SHACL validation for enforcing timing, budget, immutability, and availability constraints. The framework is evaluated offline on the OULAD dataset using snapshots constructed relative to each assessment at two decision horizons. Results show that the predictive component provides strong performance, that compact intervention plans can be generated at scale, and that semantic validation reveals infeasible plans that lighter optimization-only settings would otherwise accept. Rather than claiming causal improvement in student outcomes, this work shows that counterfactual recourse becomes more operationally meaningful in education when recommendations are not only model-valid, but also semantically feasible and machine-checkable.
\end{abstract}


\vspace{-0.5cm}
\section{Introduction}
Learning analytics has made progress in predicting student failure, disengagement, and dropout risk from educational traces, assessment records, and learner profiles. However, predictive performance alone is insufficient for effective educational decision support. In practice, instructors, advisors, and student support services need recommendations that are actionable, contextually appropriate, and operationally feasible, rather than risk scores in isolation. Recent work in learning analytics and AI in education has highlighted this need to move beyond prediction toward systems that better support intervention, human oversight, and trustworthiness~\cite{akccapinar2019using,huang2023personalized,alalawi2025evaluating}.

A promising direction for bridging prediction and action lies in counterfactual explanations and algorithmic recourse. These approaches aim to identify what should change in order to obtain a more desirable model outcome, with recourse methods emphasizing actionability for the affected individual~\cite{ustun2019actionable,mothilal2020explaining}. In educational settings, however, application of recourse remains challenging. Many methods operate primarily in feature space and may therefore generate recommendations that are mathematically valid but difficult to enact in practice. Educational recommendations must often respect limited time before an assessment, bounded intervention effort, immutable learner attributes, and the availability of pedagogical resources. In addition, explainable learning analytics has shown that usefulness depends not only on interpretability, but also on stability, practical validity, and alignment with stakeholder needs~\cite{tiukhova2024explainable}.

This paper addresses these limitations through \textbf{SC$^2$R} -- \textit{Semantics-Constrained Counterfactual Recourse} -- a framework for educational decision support that combines predictive modeling, integer-programming-based recourse generation over discrete action variables, an OWL/SKOS vocabulary for intervention actions, and SHACL validation for enforcing timing, budget, immutability, and optional availability constraints~\cite{skos2009,shacl2017}. The contribution of the paper is intentionally focused: rather than claiming causal educational impact, we investigate whether counterfactual recourse becomes more reliable and operationally meaningful when its outputs are represented as machine-checkable intervention plans subject to explicit semantic constraints. The results show that constrained optimization can generate compact plans at scale and, more importantly, that richer semantic constraints expose infeasible recommendations that lighter optimization-only settings would otherwise accept. These findings suggest that, for educational decision support, counterfactual recourse becomes substantially more defensible when recommendations are not only model-valid, but also semantically validated, interpretable, and operationally feasible.

The remainder of this paper is organized as follows. Section \ref{sec_relatedworks} reviews the relevant literature. Section \ref{sec_problem} formalizes the problem. Section \ref{sec_framework} presents \textbf{SC$^2$R}, including its predictive, recourse, and semantic validation components. Section \ref{sec_experiment_settings} describes the experimental setup and evaluation protocol. Section \ref{results} reports the empirical results. Section \ref{sec_discussion} discusses the findings and limitations. Finally, Section \ref{sec_conclusion} concludes the paper and outlines directions for future work.
	
\section{Related Work}
\label{sec_relatedworks}

The present work lies at the intersection of learning analytics for student support, counterfactual explanations and algorithmic recourse, and semantic technologies for validation and feasibility checking. Although each of these areas has advanced substantially, their integration into a unified framework for educational decision support remains limited.

Learning analytics has produced a large body of work on the prediction of student failure, disengagement, and dropout from educational traces, assessment histories, and learner profiles. These studies have demonstrated the value of predictive models for identifying students who may require support. However, several studies have also emphasized that predictive performance alone is not sufficient for educational usefulness. In practice, instructors, advisors, and support services need outputs that can inform timely, actionable, and contextually meaningful intervention, rather than risk scores in isolation~\cite{alalawi2025evaluating}. This limitation is important in educational settings, where recommendations are typically interpreted by human stakeholders and must remain compatible with pedagogical and institutional constraints. Recent work has improved the reliability of educational early-warning models through leakage-excluded evaluation~\cite{le2026can}, but has not addressed actionable recourse.

Counterfactual explanations and algorithmic recourse help bridge this gap by identifying changes that could lead to a more desirable model outcome~\cite{wachter2017counterfactual,slack2021counterfactual,ustun2019actionable}. Prior work has studied desirable properties such as sparsity, diversity, plausibility, robustness, and feasibility~\cite{mothilal2020explaining,ferrario2022robustness,sharma2024faster}, and has increasingly framed recourse as an intervention-oriented problem under actionability constraints~\cite{karimi2021algorithmic}. However, most existing approaches remain defined mainly in feature space, which can produce recommendations that are model-valid but difficult to justify or implement in practice.

This limitation is particularly important in education, where recommendations must satisfy temporal, organizational, and pedagogical constraints. Suggested actions may need to occur before an assessment, remain within a bounded intervention effort, avoid changing immutable learner attributes, and match the actual availability of pedagogical resources. Recent work in explainable learning analytics also emphasizes that usefulness depends not only on interpretability, but also on practical validity, stability, and alignment with stakeholder expectations~\cite{tiukhova2024explainable}.

Semantic Web technologies offer a natural way to formalize such requirements. SHACL provides a W3C-standard mechanism for validating RDF graphs against declarative constraints and is therefore well suited to expressing conditions related to timing, budget, immutability, and resource availability~\cite{shacl2017}. Although widely used in knowledge graph validation, such semantic technologies have rarely been integrated into counterfactual recourse pipelines.


Against this background, this paper combines predictive modeling, constrained counterfactual recourse, and semantic validation within a unified framework for educational decision support. The objective is not only to generate recommendations that are model-valid, but also to represent them as machine-checkable, interpretable, and semantically feasible intervention plans. In this respect, the paper addresses a gap left by prior work, which has generally treated prediction, recourse, and semantic validation as separate concerns rather than as components of the same decision-support pipeline.

\section{Problem Formulation}
\label{sec_problem}

We consider the problem of generating feasible intervention plans for students in a digital learning environment. Let $x_t \in \mathbb{R}^p$ denote the state of a student observed at snapshot time $t$, with snapshots constructed at predefined times relative to the due date $d$ of an upcoming assessment or learning milestone (e.g., $t=d-14$ or $t=d-7$). This state is constructed from information typically available in learning management systems, including learner profile attributes, prior assessment history, interaction traces with learning resources and activities, engagement indicators, and temporal variables derived from the course schedule. Based on this representation, a predictive model $f_h(x_t)$ estimates the probability of success on the next assessment, where $h=d-t$ denotes the remaining time-to-deadline horizon (e.g., $h=14$ or $h=7$ days).

The objective is not only to predict the likelihood of success or failure, but also to identify candidate actions that may support a more desirable outcome. To this end, we define a counterfactual intervention plan as $\pi = \{a_1, \ldots, a_m\}$, where each $a_i$ denotes a discrete, human-readable intervention action. Such actions are intended to modify aspects of learner engagement or study behavior within the remaining time window before the next assessment. Applying the plan $\pi$ transforms the student state into a post-intervention state denoted by $x_t \oplus \pi$.

The problem is then to identify an intervention plan that minimizes intervention burden while satisfying a target decision threshold:
\vspace{-0.3cm}
\begin{equation}
	\min_{\pi} \; c(\pi)
	\quad
	\text{s.t.}
	\quad
	f_h(x_t \oplus \pi) \ge \tau,\;
	\pi \in \mathcal{A},\;
	\mathcal{S}(G_\pi)=1.
\end{equation}

\vspace{-0.2cm}
Here, $\tau$ denotes a generic target decision threshold for recourse generation. The numeric value $\tau = 0.60$ shown in Fig.~\ref{fig:example} is purely illustrative and is included only to clarify how candidate plans are accepted. This formulation is consistent with the broader recourse literature, where the objective is to identify low-cost changes that achieve a desired model outcome while remaining actionable~\cite{wachter2017counterfactual,ustun2019actionable,karimi2021algorithmic}. Here, $c(\pi)$ denotes a weighted intervention cost, $\mathcal{A}$ is the allowable action space, $G_\pi$ is the RDF graph representing the intervention plan, and $\mathcal{S}(G_\pi)$ is a binary SHACL conformance function. Under this formulation, a plan is acceptable only if it both achieves the desired predictive objective and satisfies the semantic feasibility constraints encoded in the validation layer.

\begin{figure}[t]
	\centerline{\includegraphics[width=0.85\linewidth]{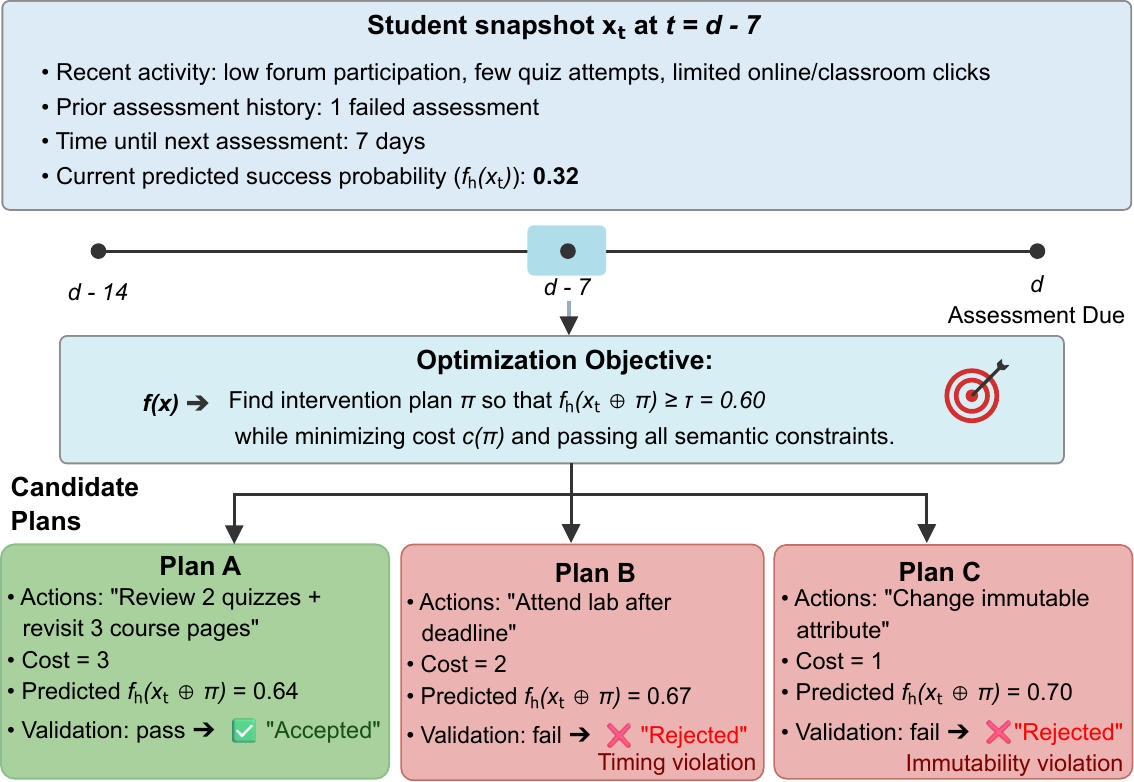}}
	\caption{Illustration of the problem formulation. Candidate plans are retained only if they satisfy both the prediction target and the semantic constraints.}
	\label{fig:example}
	\vspace{-0.7cm}
\end{figure}

The constraints considered in this work are intentionally explicit and operational. First, \emph{timing constraints} require that all actions be schedulable before the due date of the next assessment and within the available snapshot-to-deadline window. Second, \emph{budget constraints} ensure that the total intervention burden remains below a configurable upper bound. Third, \emph{immutability constraints} prevent the recourse mechanism from modifying fixed learner characteristics or non-actionable assessment metadata. Fourth, an optional \emph{availability constraint} requires that actions involving learning resources or activities be compatible with their availability conditions in the platform.

As illustrated in Fig.~\ref{fig:example}, educational recourse is therefore not treated as a purely geometric perturbation problem in feature space, but as a constrained decision-support problem. In this setting, a recommendation is useful only if it is not only model-valid, but also feasible to enact and understandable to human stakeholders. The next section describes the framework used to operationalize this formulation.

\section{SC$^2$R: Semantics-Constrained Counterfactual Recourse Framework}
\label{sec_framework}

\textbf{SC$^2$R} consists of three main components: a predictive component for estimating next-assessment success, a recourse component for generating candidate intervention plans, and a semantic validation layer for checking feasibility. Together, these components turn counterfactual recourse into a machine-checkable decision-support process.

Formally, \textbf{SC$^2$R} proceeds as follows. Starting from a student snapshot $x_t$, the predictive component computes $f_h(x_t)$, the probability of success at horizon $h$. The recourse component searches for an intervention plan $\pi \in \mathcal{A}$ that minimizes the intervention cost $c(\pi)$ while producing a post-intervention state $x_t \oplus \pi$ such that $f_h(x_t \oplus \pi) \ge \tau$. Each candidate plan is represented as an RDF graph $G_\pi$ and retained only if it satisfies the semantic feasibility condition $\mathcal{S}(G_\pi)=1$.

Fig.~\ref{fig:architecture} presents the overall \textbf{SC$^2$R} framework. Starting from data available in a learning management system or digital learning environment, the pipeline first constructs snapshot representations at predefined decision horizons. A calibrated predictive model is then applied to estimate the probability of success on the next assessment. Based on this prediction, an optimization module generates candidate intervention plans over a discrete action space. These plans are represented as RDF graphs and passed through a semantic validation layer based on SHACL. Only plans that satisfy both the predictive objective and the explicit semantic constraints are retained for downstream evaluation. The resulting outputs are finally analyzed through predictive and recourse evaluation metrics.

\begin{figure}[t]
	\centerline{\includegraphics[width=0.85\linewidth]{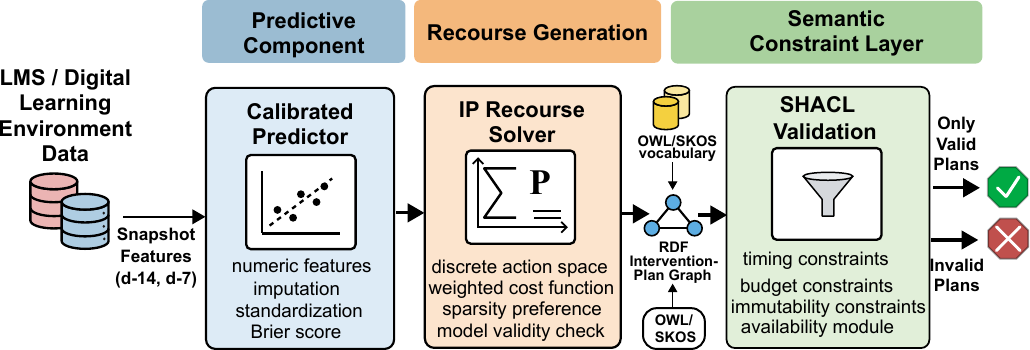}}
	\caption{Overview of \textbf{SC$^2$R}. From student snapshots, the pipeline predicts next-assessment success, generates intervention plans, applies semantic validation, and retains only conformant plans for evaluation.}
	\label{fig:architecture}
	\vspace{-0.7cm}
\end{figure}

\subsection{Predictive Component}
The predictive component maps each student snapshot $x_t$ to a probability of success $f_h(x_t)$ for the next assessment at horizon $h$. In this work, the primary predictor is a calibrated logistic-regression pipeline. This choice is deliberate. A linear model is transparent, computationally efficient, and compatible with actionable recourse formulations for linear decision boundaries~\cite{ustun2019actionable}. Numeric features are imputed and standardized before classification. Model calibration is assessed using the Brier score, which is especially relevant because the downstream recourse mechanism operates on predicted probabilities rather than on hard class labels~\cite{glenn1950verification,yang2022modified}.

Although the broader implementation also includes stronger nonlinear models, such as gradient-boosted trees and neural architectures, the main paper does not rely on them. Their role is supplementary, whereas the calibrated linear baseline provides a more interpretable foundation for the \textbf{SC$^2$R} framework developed here.

\subsection{Recourse Generation}
Given a student state $x_t$ and predictor $f_h$, the recourse component searches over the allowable action space $\mathcal{A}$ for an intervention plan $\pi$ that minimizes the weighted cost $c(\pi)$ while ensuring that the post-intervention state $x_t \oplus \pi$ satisfies the target condition $f_h(x_t \oplus \pi) \ge \tau$. Rather than recommending arbitrary perturbations in feature space, the optimizer selects among interpretable intervention increments derived from learner engagement and study-related variables, such as increasing interaction with particular categories of learning resources or activities during the remaining study window. The total intervention burden is modeled through a weighted cost function, optionally combined with a sparsity preference so that shorter plans are favored.

The main solver is formulated as an integer program. This choice is natural because intervention variables are discrete, several feasibility conditions are combinatorial, and the optimization objective explicitly minimizes intervention burden. A generated plan is considered \emph{model-valid} if the corresponding post-intervention state crosses the target decision threshold under the predictive model, that is, if $f_h(x_t \oplus \pi) \ge \tau$. However, model validity alone is not sufficient. In implementation, the optimization module first generates candidate plans over the discrete action space, after which semantic feasibility is enforced through SHACL validation. Thus, the formal condition $\mathcal{S}(G_\pi)=1$ is realized as a validation step between plan generation and downstream evaluation.

\subsection{Semantic Validation Layer}
The semantic validation layer provides the representation and checking mechanisms that make candidate intervention plans operationally meaningful. It combines a lightweight ontology for intervention-plan representation with SHACL-based validation constraints.


We define a lightweight RDF vocabulary for intervention-plan representation, using the prefix \texttt{cbe:} for terms introduced within a competency-based education ontology extended here for recourse representation~\cite{le2025vers}. The vocabulary is centered on two main classes, \texttt{cbe:ActionPlan} and \texttt{cbe:Action}. An intervention plan is modeled as an instance of \texttt{cbe:ActionPlan}, linked to one or more actions through \texttt{cbe:hasAction}, and associated with the target student and target assessment through \texttt{cbe:forStudent} and \texttt{cbe:forAssessment}, as shown in Fig.~\ref{fig:ontology}. At the data level, a plan records the snapshot day, due day, and budget, while each action records its activity type, intended engagement change, action cost, and scheduled time interval. We combine a lightweight ontology with SKOS because these two layers serve complementary purposes: the ontology captures the formal structure of plans, actions, students, and assessments, whereas SKOS provides a controlled and extensible classification of activity types. For example, an intervention step can be represented structurally as an instance of \texttt{cbe:Action} linked to a \texttt{cbe:ActionPlan}, while its pedagogical category is assigned through a SKOS concept such as quiz practice, forum participation, or resource review. This design makes the action space explicit, reusable across plans, and easy to refine or align without introducing unnecessary ontological commitments.

\begin{figure}[t]
	\centerline{\includegraphics[width=0.85\linewidth]{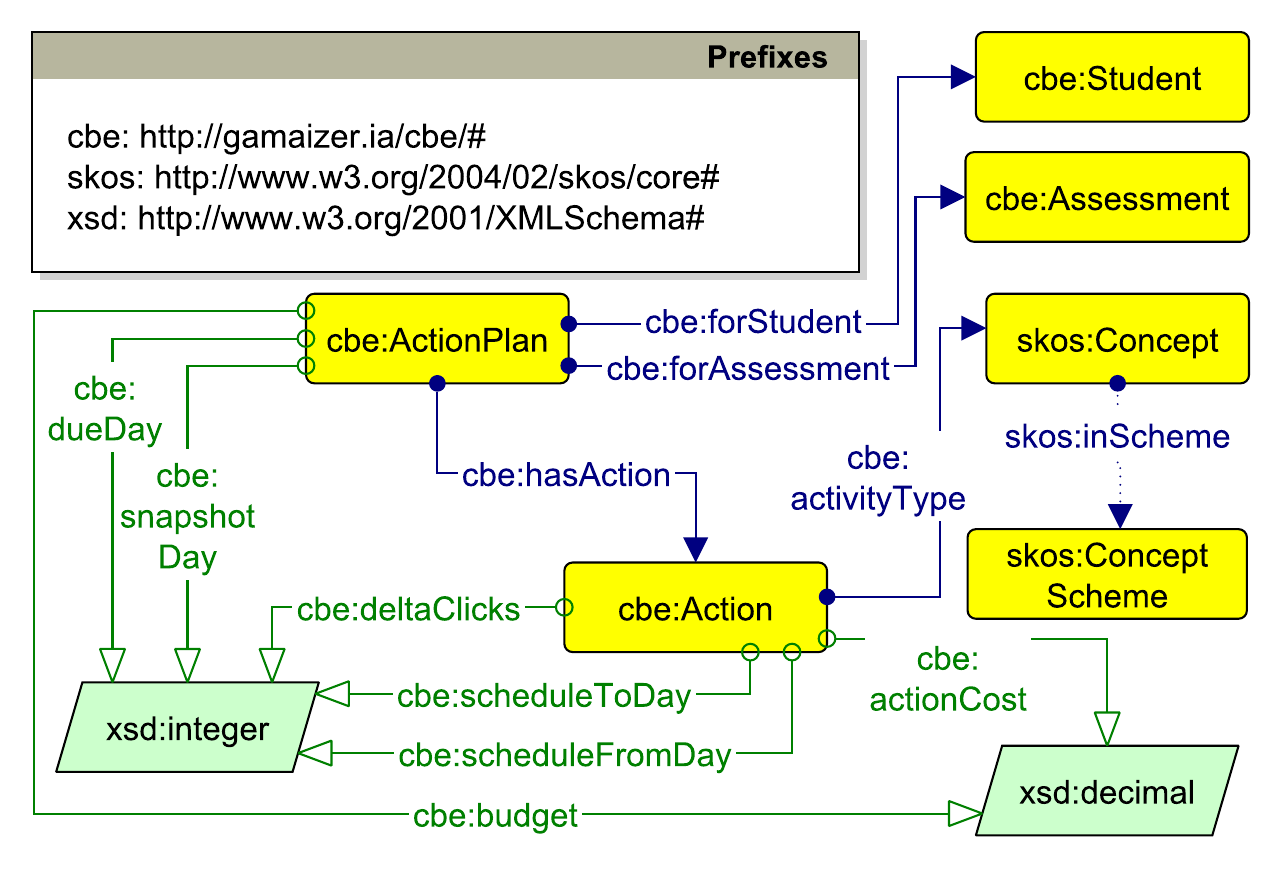}}
	
	\vspace{-0.3cm}
	\caption{Ontology fragment for representing intervention plans, centered on \texttt{cbe:ActionPlan} and \texttt{cbe:Action}.}
	
	\label{fig:ontology}
	\vspace{-0.7cm}
\end{figure}

\vspace{-0.4cm}
\begin{lstlisting}[caption={Compact SHACL summary.},label={lst:shacl-compact},basicstyle=\ttfamily\scriptsize]
ActionPlanShape:
 targetClass ActionPlan
 requires {snapshotDay:int, dueDay:int, budget:decimal,
         hasAction -> ActionShape}	
ActionShape:
 targetClass Action
 requires {activityType, deltaClicks:int>=0, 
         actionCost:decimal>=0,
         scheduledFromDay:int, scheduledToDay:int}
SPARQL constraints:
 C1 (timing):        snapshotDay <= from <= to <= dueDay
 C2 (budget):        sum(actionCost) <= budget
 C3 (availability):  availableFromDay(activityType) <= from
       and to <= availableToDay(activityType)
\end{lstlisting}
\vspace{-0.2cm}

Each candidate intervention plan $\pi$ generated by the recourse module is encoded as an RDF graph $G_{\pi}$ using this vocabulary. This representation makes the plan machine-readable and structured, while preserving the distinction between the plan as a whole and the individual actions it contains.

SHACL shapes are used to determine whether $\mathcal{S}(G_{\pi})=1$, that is, whether the RDF graph of the candidate plan satisfies the semantic feasibility conditions of \textbf{SC$^2$R}~\cite{shacl2017}. At the structural level, the shapes require the presence of the core plan and action properties. At the constraint level, they enforce timing and budget consistency, and optionally availability compatibility for activity types whose valid time windows are known. Listing~\ref{lst:shacl-compact} summarizes the main shapes and constraints used in the current implementation.

Table~\ref{tab:action_space} summarizes the intervention vocabulary used in \textbf{SC$^2$R}, including the affected feature(s), admissible increments, intervention costs, and semantic feasibility conditions. It clarifies how candidate recourse actions are grounded in interpretable educational interventions and subsequently validated through the semantic layer.

\begin{table}[t]
	\centering
	\caption{Compact intervention vocabulary used in \textbf{SC$^2$R}.}
	\vspace{-0.2cm}
	\label{tab:action_space}
	\scriptsize
	\setlength{\tabcolsep}{3pt}
	\begin{tabular}{p{1.6cm}p{1.4cm}p{1.0cm}p{0.5cm}p{3.3cm}}
		\toprule
		\textbf{Intervention} & \textbf{Feature} & \textbf{Increments} & \textbf{Cost} & \textbf{Constraints} \\
		\midrule
		Review quizzes & Quiz clicks & +1,+2,+3 & 1/inc. & Before due day; available \\
		Revisit pages & Page clicks & +1 to +5 & 1/2p & Before due day \\
		Attend support & Support count & +1 & 3 & Session available; before due day \\
		Prep activity & Prep feature & +1 & 2 & Schedule-compatible \\
		\bottomrule
	\end{tabular}
	\vspace{-0.6cm}
\end{table}

In addition to these checks, the semantic layer supports the immutability principle adopted in this work by ensuring that candidate plans act only on allowable intervention variables and not on fixed learner characteristics or non-actionable assessment metadata. This layer is essential because feasibility cannot be reduced to geometric proximity in feature space. Two plans may have similar numerical cost while differing substantially in whether they can be completed before the assessment deadline or whether the required pedagogical resources are actually available. The semantic layer therefore provides an explicit feasibility-checking layer around the recourse optimizer, ensuring that retained recommendations are not only model-valid, but also feasible and interpretable.

Taken together, \textbf{SC$^2$R} provides a framework in which predictive modeling, constrained recourse generation, and semantic validation are treated as parts of the same decision-support pipeline. The result is not merely a set of counterfactual feature changes, but a structured intervention artifact that can be checked and interpreted before any downstream use.
\section{Experimental Setup}
\label{sec_experiment_settings}

This section presents the dataset, snapshot protocol, comparison methods, and evaluation metrics used in the study\footnote{The implementation is publicly available at  \url{https://github.com/lengocluyen/semantics-constrained-counterfactuel-recourse}.}.

\subsection{Dataset and Snapshot Protocol}
We evaluate \textbf{SC$^2$R} on the OULAD dataset, which provides student demographics, assessment schedules and scores, virtual learning environment (VLE) metadata, and daily interaction summaries across multiple module presentations~\cite{kuzilek2017open}. We use the tables \texttt{studentInfo}, \texttt{assessments}, \texttt{studentAssessment}, \texttt{vle}, and \texttt{studentVle}.

For each assessment with due day $d$, we construct two snapshots per eligible student, at $t=d-14$ and $t=d-7$. Only information available up to the snapshot time is used; post-$t$ interactions and assessments are excluded. Early assessments that would otherwise induce negative snapshot times are handled during preprocessing.

The feature set includes aggregated click counts by activity type over rolling 7-, 14-, and 28-day windows, availability-alignment features, prior assessment history, time-to-deadline variables, and immutable learner descriptors. The prediction target is next-assessment pass/fail, where pass is defined as score $\ge 40$.
To avoid leakage across closely related offerings, data are split at the module-presentation level. The frozen split contains 15 training presentations, 2 validation presentations, and 5 test presentations. Unless otherwise stated, results are reported on the held-out test set, with $d\!-\!14$ as the main horizon and $d\!-\!7$ used as a secondary horizon.

\begin{table}[t]
	\caption{Dataset and protocol summary.}
	\vspace{-0.2cm}
	\label{tab:data}
	\centering
	\renewcommand{\arraystretch}{1.0}
	\begin{tabular}{p{0.38\columnwidth} p{0.52\columnwidth}}
		\toprule
		\textbf{Item} & \textbf{Value} \\
		\midrule
		Snapshot horizons & $d\!-\!14$, $d\!-\!7$ before next assessment \\
		Train / val / test presentations & $15 / 2 / 5$ \\
		Instances per horizon (train / val / test) & $190{,}752 / 29{,}054 / 84{,}791$ \\
		Primary label & Next-assessment pass/fail ($\mathrm{score}\geq 40$) \\
		Primary recourse method & Integer programming on action variables \\
		Full-scale IP plans scored offline & $127{,}972$ \\
		\bottomrule
	\end{tabular}
	\vspace{-0.7cm}
\end{table}

\subsection{Comparison Methods}
The primary predictive component is a calibrated logistic-regression model, used as the reference predictor for \textbf{SC$^2$R}. Although the implementation also includes stronger nonlinear models, including XGBoost~\cite{chen2016xgboost}, a TabTransformer-style architecture~\cite{huang2020tabtransformer}, and a BiLSTM-based model~\cite{hochreiter1997long,schuster1997bidirectional}, these are treated as supplementary analyses.

For recourse generation, the main method is an integer-programming (IP) formulation over discrete action variables. As a baseline, we use a Wachter-style counterfactual search~\cite{wachter2017counterfactual} on a controlled 200-case subset. A stricter IP variant with availability constraints is also included to assess the added value of richer semantic validation. 

\subsection{Evaluation Metrics}
We evaluate the framework along two dimensions: (i) predictive quality is measured using AUC, F1, accuracy, average precision, and Brier score; and (ii) recourse quality is measured using model validity, SHACL conformance, intervention cost, number of actions, predicted probability gain, $k$-nearest-neighbor plausibility, and stability under noise and retraining.

\section{Experimental Results}
\label{results}

\subsection{Predictive Quality}
Fig.~\ref{fig:predictive_overview} summarizes the predictive results on the test set. The calibrated logistic-regression baseline provides strong performance, with AUC values of $0.884$ at $d\!-\!14$ and $0.889$ at $d\!-\!7$. F1 and average precision follow the same pattern, indicating that the later snapshot benefits from more recent learner evidence. The Brier scores are $0.128$ and $0.122$, which indicate usable calibration for downstream recourse generation.

Supplementary checks with stronger nonlinear models do not materially change the central picture of the paper. On the $d\!-\!14$ split, XGBoost and the TabTransformer-style model reach an AUC of $0.898$, while the MLP reaches $0.895$ and the BiLSTM $0.888$. These improvements remain modest relative to the calibrated logistic baseline. For this reason, logistic regression remains the reference predictor for \textbf{SC$^2$R}.

\begin{figure}[t]
	\centering
	\includegraphics[width=0.99\linewidth]{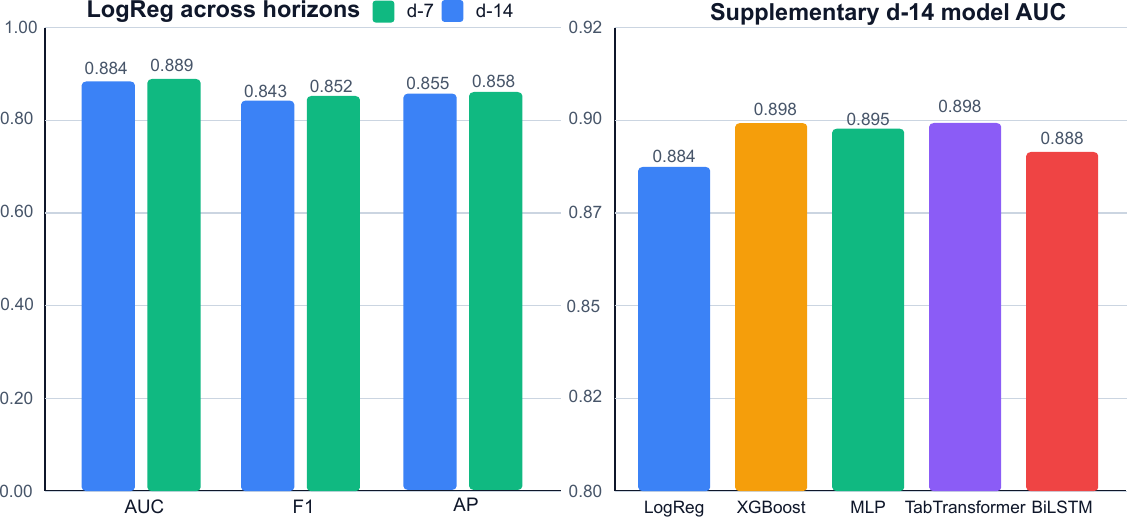}
	\caption{Predictive performance overview. Left: logistic regression performs strongly at both decision horizons, with slightly better results at $d\!-\!7$. Right: supplementary nonlinear models yield only modest AUC gains on the $d\!-\!14$ split.}
	\label{fig:predictive_overview}
	\vspace{-0.3cm}
\end{figure}

\subsection{Recourse Quality at Scale}
The main full-scale integer-programming evaluation produces $127{,}972$ scored plans. Table~\ref{tab:recourse} summarizes the results. Both model validity and SHACL conformance reach $1.0$. The generated plans are also compact, with an average of $1.075$ actions and a mean total cost of $10.357$. The mean predicted probability gain is $0.383$. These results show that \textbf{SC$^2$R} can generate low-cardinality intervention plans at scale while satisfying both the predictive objective and the semantic feasibility constraints considered in the main evaluation setting.

Plausibility provides a more nuanced picture. The mean 5-nearest-neighbor distance of the recourse-adjusted states is $6.950$, suggesting that the generated plans remain close to observed learner profiles without collapsing to trivial copies of successful historical cases. Stability under noise and retraining is discussed together with the semantic ablation in Fig.~\ref{fig:semantic_ablation}.


\begin{table}[t]
	\caption{Full-scale IP recourse summary.}
	\vspace{-0.2cm}
	\label{tab:recourse}
	\centering
		\begin{tabular}{p{3.0cm}p{0.6cm}p{3.0cm}p{0.6cm}}
			\toprule
			Metric & Value & Metric & Value \\
			\midrule
			Plans scored & $127{,}972$ & Mean number of actions & 1.075\\
			SHACL conformance rate & 1.000 &Mean predicted gain & 0.383 \\
			Model validity rate & 1.000 & Mean 5-NN plausibility distance & 6.950 \\
			Mean total cost & 10.357 & 	Mean noise stability & 0.656\\
			Median total cost & 11.000 &Mean retrain stability & 0.228 \\
			\bottomrule
	\end{tabular}
	\vspace{-0.7cm}
\end{table}

\subsection{Effect of Semantic Validation}
The contribution of the semantic layer is most visible in the controlled 200-case ablation shown in Fig.~\ref{fig:semantic_ablation}. The IP solver and the Wachter-style baseline have nearly identical average cost (24.127 vs.\ 24.175), and both achieve perfect model validity and SHACL conformance under the baseline constraint configuration. However, when availability constraints are activated in the IP pipeline, the conformance rate decreases to $0.875$ without any reduction in optimization cost. This is an informative result rather than a negative one: it shows that richer semantic constraints reveal plans that would otherwise appear acceptable under a lighter constraint regime.

Fig.~\ref{fig:semantic_ablation} also reports the robustness indicators associated with the generated plans. Noise stability is moderate ($0.656$), whereas retrain stability is lower ($0.228$). This suggests that the recourse outputs are more stable under small perturbations of the input than under changes in the learned decision boundary.

\begin{figure}[t]
	\centering
	\includegraphics[width=0.85\linewidth]{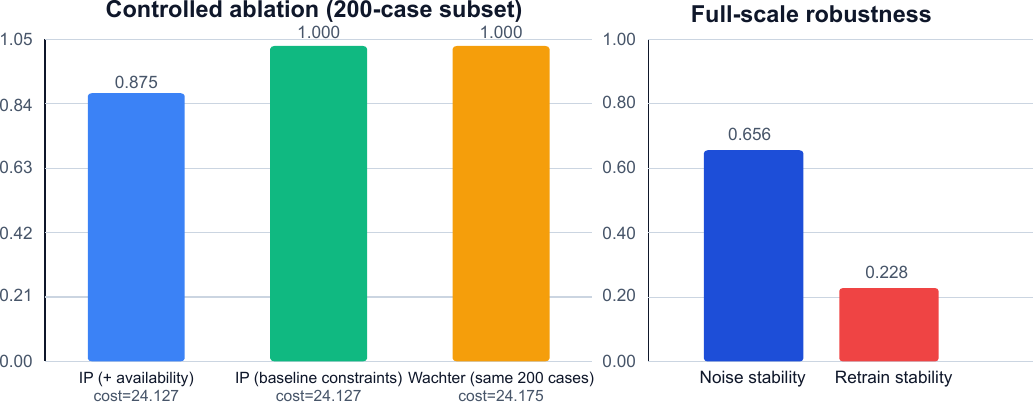}
	\caption{Effect of semantic validation and robustness. Left: 200-case ablation comparing IP recourse, Wachter-style recourse, and IP recourse with availability constraints. Right: robustness indicators under input noise and retraining.}
	\label{fig:semantic_ablation}
	\vspace{-0.6cm}
\end{figure}

\section{Discussion}
\label{sec_discussion}

The results support a focused interpretation of \textbf{SC$^2$R}. First, the predictive component provides a strong and well-calibrated basis for downstream recourse generation. Second, the integer-programming formulation produces compact intervention plans at scale, suggesting that discrete recourse can be operationalized efficiently in this setting. Third, and most importantly, the semantic layer changes how recourse outputs should be interpreted: its contribution is not to reduce optimization cost, but to distinguish between plans that are merely model-valid and plans that remain feasible under explicit timing, budget, immutability, and availability constraints.

At the same time, the scope of the study remains intentionally limited. The evaluation is offline, model-based, and observational, and therefore does not establish that following a generated plan would improve student outcomes. In addition, the current action vocabulary is necessarily simplified and derived from structured learner activity signals rather than richer pedagogical interaction data. The lower retrain stability also suggests that some recommendations may remain sensitive to changes in the learned predictive boundary.

These observations indicate that \textbf{SC$^2$R} should be understood as a framework for machine-checkable and semantically feasible educational recourse, rather than as evidence of causal intervention effectiveness. Its value lies in structuring and validating candidate intervention plans before any practical use by instructors, advisors, or learning-support services.

\section{Conclusion and Perspective}
\label{sec_conclusion}

This paper introduced \textbf{SC$^2$R}, a semantics-constrained counterfactual recourse framework for educational decision support. Using the OULAD dataset, the results showed that a calibrated logistic-regression model provides a strong reference predictor, that integer-programming-based recourse can generate compact plans at scale, and that semantic validation plays a central role in detecting hidden infeasibility beyond optimization alone.
The main perspective of this work is to move from model-consistent recourse toward more educationally grounded intervention support settings. Future work will examine dynamic replanning, institution-specific constraints, and recourse robustness across retrained models.

\section*{Acknowledgments}
We warmly thank the Ikigai consortium led by the association Games for Citizens, the company Gamaizer, as well as the FORTEIM project (winner of the AMI CMA France 2030 call for projects), for their support and collaboration. Their contributions have provided significant added value to the completion of this research.

\bibliographystyle{ieeetr}
\bibliography{references}

\end{document}